\documentclass[USletter,11pt]{article}

\usepackage{changepage}

\input{setup.sty}

\title{The Symmetrized Holographic Entropy Cone}
 
\author[a]{Matteo Fadel}
\emailAdd{fadelm@phys.ethz.ch}

\author[b]{\!\!, Sergio Hern\'andez-Cuenca}
\emailAdd{sergiohc@ucsb.edu}

\affiliation[a]{Department of Physics, ETH Z\"{u}rich, 8093 Z\"{u}rich, Switzerland}

\affiliation[b]{Department of Physics, University of California, Santa Barbara, CA 93106, USA}

\abstract{The holographic entropy cone (HEC) characterizes the entanglement structure of quantum states which admit geometric bulk duals in holography. Due to its intrinsic complexity, to date it has only been possible to completely characterize the HEC for at most $n=5$ numbers of parties. For larger $n$, our knowledge of the HEC falls short of incomplete: almost nothing is known about its extremal elements. Here, we introduce a symmetrization procedure that projects the HEC onto a natural lower dimensional subspace. Upon symmetrization, we are able to deduce properties that its extremal structure exhibits for general $n$. Further, by applying this symmetrization to the quantum entropy cone, we are able to quantify the typicality of holographic entropies, which we find to be exponentially rare quantum entropies in the number of parties.
}

\begin{document}
 
 
\maketitle
 
\clearpage
\newpage

\section{Introduction}\label{sec:intro}

From the microscopic degrees of freedom of gravity to the macroscopic emergence of spacetime itself, the holographic principle has provided deep insight into many aspects of quantum gravity~\cite{tHooft:1993dmi,Susskind:1994vu}. Realizations thereof like AdS/CFT~\cite{Maldacena_1999,Witten1998} may be used to define a bulk theory of quantum gravity non-perturbatively in terms of a lower-dimensional boundary field theory -- a hologram. Elucidating quantum gravity then amounts to decoding the holographic dictionary between bulk and boundary physics. A remarkably powerful entry in this dictionary concerns the correspondence between boundary entanglement and bulk geometry, and is often captured by the slogan that spacetime emerges from entanglement~\cite{VanRaamsdonk:2010pw}.

At the heart of this idea lies the Ryu-Takayanagi (RT) formula~\cite{Ryu:2006bv} (see~\cite{Hubeny:2007xt,Faulkner:2013ana,Engelhardt:2014gca} for generalizations thereof), which geometrizes the von Neumann entropy of boundary regions into areas of certain surfaces in the bulk. By RT, the emergence of a classical spacetime from entanglement relies on strong constraints on the entanglement structure of boundary states: only so-called holographic states with certain patterns of quantum entanglement admit such geometrizations. When formulated as inequalities for the von Neumann entropy, these constraints define a polytope in the space of entropies of all subsystems one can form involving $n$ parties. In~\cite{Bao:2015bfa}, this object was proven to be a rational polyhedral cone for any given $n$ and coined the holographic entropy cone (HEC). 
The work of~\cite{Bao:2015bfa} also pioneered a graph-theoretic reformulation of RT entropies, later formalized by~\cite{Avis:2021xnz}, that made the HEC amenable to systematic study. In particular,~\cite{Bao:2015bfa} proved that the HEC could be equivalently defined as the space of entropies that can be realized by the weights of minimum cuts on weighted graphs. Despite this breakthrough, deciphering the general structure of the HEC$_n$ for general $n$ remains a formidable challenge: its complete description is only known for up to $n=5$ parties~\cite{HernandezCuenca:2019wgh}, and despite the systematic computational algorithms of~\cite{Avis:2021xnz} for arbitrary $n$, results for $n\ge6$ are only partial and highly incomplete~\cite{Bao:2015bfa,n6wip}.

The remarkably complicated structure the HEC$_n$ exhibits as $n$ increases suggests that an explicit description of all its rich details, if attainable at all, would probably not be very illuminating. Hence, rather than trying to explicitly describe the HEC in full glory, some attempts at studying it have focused on trying to understand its basic structural properties~\cite{Hubeny:2018trv,Hubeny:2018ijt,Hernandez-Cuenca:2019jpv,He:2019ttu,He:2020xuo}. In pursuing the latter strategy, the present work identifies an interesting projection of the HEC which is able to wash out its uninteresting fine details while preserving non-trivial features of the extremal structure of the HEC$_n$ at arbitrary $n$. Since this projection is inspired by the underlying symmetries of the HEC, we will be referring to it as a symmetrization. Similar approaches have been extremely successful in the context of Bell nonlocality, where a projection of the local polytope onto the subspace invariant under permutation of the particles allowed to find Bell inequalities valid for arbitrary $n$ \cite{TuraSci,FadelPRL}. Importantly, our symmetrization of the HEC seems to take a simple enough form that we are able to propose a general solution to it, and thereby conjecture non-trivial properties that all extreme rays and facets of the HEC$_n$ must satisfy for all $n$. 

We begin in section \ref{sec:basics} with a brief review of basic quantum and holographic inequalities that will be appear throughout, and a self-contained introduction to the graph models of holographic entanglement of \cite{Bao:2015bfa}. Having laid out our notation, in Section \ref{sec:symm} we then carefully explain the symmetrization operations that are central to our explorations. Section \ref{sec:shec} presents our explicit results and conjectures for the general form of the HEC upon symmetrizations. By looking at analogous general-$n$ results for arbitrary quantum states in Section \ref{sec:sqec}, we are then able to quantify the typicality of holographic entropies in Section \ref{sec:vols}: with respect to the uniform volume measure of entropy space, the latter are exponentially rare in the number of parties. We conclude in Section \ref{sec:end} with a summary of results and a discussion of open questions.

\section{Basics}\label{sec:basics}

\subsection{Inequalities}\label{sec:ineqs}

Let us collect here some basic inequalities which will be useful for later reference. The simplest universal quantum inequality involves $2$ parties and is known as subadditivity (SA). Proven by \cite{lanford1968mean}, for arbitrary disjoint subsystems $I$ and $J$, SA reads
\begin{equation}
\label{eq:sa}
    S_I + S_J \geq S_{I\cup J} \;.
\end{equation}
This inequality holds trivially in holography. At $3$ parties, there appears another well-known universal quantum inequality: strong subadditivity (SSA). Proven by \cite{lieb1973proof}, SSA for arbitrary overlapping subsystems $I$ and $J$ reads\footnote{To clarify, this is a $3$-party inequality because it involves $3$ disjoint subsystems: $I\smallsetminus J$, $J\smallsetminus I$, and $I\cap J$.}
\begin{equation}
\label{eq:ssa}
    S_{I} + S_{J} \geq S_{I\cap J} + S_{I\cup J} \;.
\end{equation}
The proof of this inequality in holography is non-trivial and was established by \cite{Headrick:2007km,Wall:2012uf}.
The first genuinely holographic inequality (i.e., valid in holography, but not in quantum theory) appears for $3$ parties and is known as the monogamy of mutual information (MMI). Proven by \cite{Hayden:2011ag,Wall:2012uf}, MMI for arbitrary disjoint subsystems $I$, $J$, and $K$ reads
\begin{equation}
\label{eq:mmi}
    S_{I\cup J} + S_{I\cup K} + S_{J\cup K} \geq S_{I} + S_{J} + S_{K} + S_{I\cup J\cup K} \;.
\end{equation}


\subsection{Graphs}\label{sec:mincuts}

For any positive integer $k$, introduce the notation $[k]\equiv\{1,\dots,k\}$. Let $G=(V,E)$ be an undirected graph with finite vertex set $V$ and edge set $E\subset V\times V$ consisting of unordered pairs $(i,j)\in E$ with $i\ne j$. We make $G$ into a weighted graph with nonnegative capacities by endowing it with a weight vector $w\in\mathbb{R}_{\ge0}^{\abs{E}}$ encoding a map $E \to \mathbb{R}_{\ge0}$. Select some vertex subset $\partial V \subseteq V$ and call them \textit{boundary vertices}. One then defines a coloring as a surjective map $b:\partial V \to [n+1]$. 
The elements $i\in[n+1]$ are called \textit{parties}, and the non-empty subsets $I\subseteq [n+1]$ are called \textit{subsystems}. Altogether, this structure defines a graph model of holographic entanglement.
 
Any subset $W\subseteq V$ characterizes a \textit{cut} of $G$, defined by a set of cut edges $C(W)\subseteq E$ as
\begin{equation}\label{eq:cutedge}
    C(W) = \left\{(v, v') \in E \st v \in W ,\, v' \in W^\complement\right\}.
\end{equation}
The \textit{cut weight} is defined as the total weight of its edges $\norm{C(W)} = \sum_{e \in C(W)} w_e$. A set $W\subseteq V$ is a cut for a subsystem $I$ if it contains precisely the boundary vertices colored by $I$, i.e., if $W\cap\partial V = b^{-1}(I)$. Any cut $W$ for $I$ with minimum cut weight $\norm{C(W)}$ defines a \textit{min-cut} for $I$. The min-cut weight for a subsystem $I$ defines its \textit{entropy} $S_I$ via
\begin{equation}\label{eq:sdiscrete}
    S_I = \min \, \left\{\norm{C(W)} \st W\cap\partial V = b^{-1}(I) \right\}.
\end{equation}
In the context of holography, the above can indeed be understood as computing the von Neumann entropy of a subsystem $I$ of some pure holographic state on $[n+1]$ \cite{Bao:2015bfa}. Such a pure state can be used to encode an arbitrary $n$-party, mixed holographic state on $[n]$. Due to its quantum mechanical role, the special party $n+1$ is thus referred to as the \textit{purifier}. Notice in particular that \eqref{eq:sdiscrete} indeed reproduces the desired purification property that $S_I = S_{I^\complement}$ for complementary subsystems $I$ and $I^\complement = [n+1]\smallsetminus I$. For this reason, the \textit{entropy vector} of the full graph model is defined excluding the purifier by
\begin{equation}
\label{eq:Svec}
    S = \{ S_I \,:\, \varnothing \neq I \subseteq [n]\},
\end{equation}
where the conventional choice of ordering is by cardinality first, and then lexicographically. The set of all entropy vectors $S\in\mathbb{R}^{2^n-1}$ obtained this way defines the \textit{holographic entropy cone} (HEC) for $n$ parties, or HEC$_n$.

More generally, the definition of an entropy vector in \eqref{eq:Svec} applies to any $n$-partite quantum system, with $S_I$ denoting the von Neumann entropy of a subsystem $I\subseteq[n]$. If instead of holographic states one considers completely arbitrary quantum states, the resulting set of all $n$-party entropy vectors defines the \textit{quantum entropy cone} (QEC) for $n$ parties, or QEC$_n$ \cite{pippenger2003inequalities}. Both the HEC and the QEC are convex cones,\footnote{To be precise, it is actually only the topological closure of the latter which is a convex cone~\cite{pippenger2003inequalities}. This technicality will not be important for us.} and the former is additionally known to be polyhedral \cite{Bao:2015bfa}. Consistent with the fact that holographic states define a special class of quantum states, the HEC is a subcone of the QEC \cite{Nezami:2016zni}. In the following, we will be interested in studying the HEC and how it relates to the QEC.

\section{Symmetrizations}\label{sec:symm}

Both entropy cones of interest in this paper are clearly symmetric under permutations of the colors $[n]$. To see a larger symmetry, recall that subsystems $I\subseteq[n+1]$ obey $S_I = S_{I^\complement}$. Hence these entropy cones are symmetric not only under permutations of $[n]$, but also under the extended symmetric group $Sym_{n+1}$ of permutations of $[n+1]$ involving the purifier $n+1$. 

We would like to simplify the structure of our entropy cones by defining a \textit{symmetrization}, i.e., an operation $P$ on their elements that is invariant under the action of $Sym_{n+1}$. More explicitly, if $x$ and $y$ are two elements of an entropy cone in the same symmetry orbit, then they should have the same symmetrized form $P(x)=P(y)$. By asking that this symmetrization be a linear map, we will be guaranteed that the symmetrization of a full convex cone remain so. The symmetrized versions of the HEC and QEC will respectively be referred to as SHEC and SQEC.

For an entropy vector $S\in\mathbb{R}^{2^n-1}$, our symmetrization will be some linear map $P:S\mapsto\tilde{S}$. To build it, first note that by the $Sym_n$ symmetry of permutations of $[n]$, $P$ can only depend on the cardinality $\abs{I}$ of the coordinate $I\subseteq[n]$. Additionally, by the full action of $Sym_{n+1}$ on $[n+1]$ with the identification $S_{I^\complement}=S_{I}$, one has that coordinates of cardinalities $\abs{I}$ and $|I^\complement| = n+1-\abs{I}$ are also related. Altogether, this means that $P$ should just depend on at most $\ceil{n/2}$ different variables, associated to the distinct possible cardinalities of subsets of $[n+1]$ with complements identified. The codomain of $P$ is thus $\mathbb{R}^{\ceil{n/2}}$ and one can label the \textit{symmetric variables} $\tilde{S}_k$ of the \textit{symmetrized vector} $\tilde{S}\in\mathbb{R}^{\ceil{n/2}}$ by cardinalities $1\le k \le \ceil{n/2}$.

A natural way of defining $P$ thus employs subsets of $[n+1]$ of fixed cardinality $k\in \left[\ceil{n/2}\right]$,
\begin{equation}
    Q_n(k) = \{I\subseteq[n+1] \,:\, \abs{I}=k\}.
\end{equation}
Then the symmetric variables can be obtained by just summing over coordinates in $Q_n(k)$,
\begin{equation}
\label{eq:symn}
    \tilde{S}_k = \frac{1}{\abs{Q_n(k)}} \sum_{I\in Q_n(k)} S_I, \qquad \abs{Q_n(k)} = \binom{n+1}{k},
\end{equation}
where we have introduced a natural normalization factor accounting for the number of terms in the sum. For instance, for $n=3$, the symmetric variables in \eqref{eq:symn} are
\begin{equation}
\begin{aligned}
    \tilde{S}_1 &= \frac{1}{4} (\S{1}+\S{2}+\S{3}+\S{4}) = \frac{1}{4} (\S{1}+\S{2}+\S{3}+\S{123}), \\
    \tilde{S}_2 &= \frac{1}{6} (\S{12}+\S{13}+\S{14}+\S{23}+\S{24}+\S{34}) = \frac{1}{3} (\S{12}+\S{13}+\S{23}),
\end{aligned}
\end{equation}
where in the last equalities we used $S_{I^\complement}=S_{I}$. Crucially, notice that these sets $Q_n(k)$ automatically take care of coordinates $I\subseteq [n]$ of cardinality $\abs{I}=n+1-k$, since $I^\complement\in Q_n(k)$. Hence the operation $P : \mathbb{R}^{2^n-1} \to \mathbb{R}^{\ceil{n/2}}$ defined this way is a valid linear symmetrization invariant under $Sym_{n+1}$. For any number of parties $n$, the SHEC and SQEC will thus be convex cones defined by the sets of all symmetrized entropy vectors $\tilde{S}\in\mathbb{R}^{\ceil{n/2}}$ obtained from symmetrizations of their parent entropy vectors $S\in\mathbb{R}^{2^n-1}$ of the HEC and QEC, respectively.

In the case of the HEC, the symmetrization of entropy vectors that we just described can also be easily arrived at from a symmetrization of graph models themselves. Indeed, since weighted graphs are ultimately the object which defines the HEC, one may consider performing a symmetrization already at this level. Given a graph model $G$ for $n$ parties, consider all possible $(n+1)!$ graphs $G^\sigma$ (some of which may be equal) obtained by permutations $\sigma\in Sym_{n+1}$ of the colors $[n+1]$. If $S$ and $S^\sigma$ are respectively the entropy vectors from $G$ and $G^\sigma$, one obviously has $S_I = S^\sigma_{\sigma(I)}$, where $\sigma(I)=\{\sigma(i) \,:\, i\in I\}$, or, equivalently, $S^\sigma_I = S_{\sigma^{-1}(I)}$. Suppose one now combines two permuted graphs in a disjoint manner or glued together by identifying all vertices of the same color in both. Either way, it is straightforward to see that the resulting graph $G^\sigma \oplus G^{\sigma'}$ yields $S^\sigma+S^{\sigma'}$ as its entropy vector. With a suggestive choice of notation, let's then define a permutation-averaged graph $\tilde{G}$ by
\begin{equation}
    \tilde{G} = \frac{1}{(n+1)!} \bigoplus_{\sigma\in Sym_{n+1}} G^\sigma,
\end{equation}
where the multiplication shall be understood as rescaling the weights of the graphs. The computation of the entropy of $I$ in this graph $\tilde{G}$ gives
\begin{equation}
    \frac{1}{(n+1)!} \sum_{\sigma\in Sym_{n+1}} S^\sigma_I = \frac{k!(n+1-k)!}{(n+1)!} \sum_{J\in Q_n(k)} S_J = \tilde{S}_k, \qquad k = \min\{\abs{I},n+1-\abs{I}\},
\end{equation}
which recovers precisely the symmetric variables in \eqref{eq:symn}. In other words, the entropies of $\tilde{G}$ only depend on the cardinality of the subsystem $I$ and are given by the entries of the symmetric entropy vector $\tilde{S}$ of the original graph $G$. In this sense, we see that given a graph $G$, the operations of symmetrization and computation of entropies commute.

Having dealt with entropy vectors, we would now like to have an analogous operation that we can directly apply to the vectors $q\in \mathbb{R}^{2^n-1}$ which define entropy inequalities $qS\ge0$. To do so, consider constructing valid inequalities that we can write down using only the symmetric variables in \eqref{eq:symn}. This can be accomplished by, given an inequality $qS\ge0$, adding up all inequalities in its symmetry orbit so as to form a new inequality $q'S\ge0$ which will be valid by convexity. The coefficients of the new vector $q'$ can be easily computed to be
\begin{equation}
\label{eq:orbitsum}
    q'_I = k!(n+1-k)! \sum_{J\in Q_n(k)} q_J, \qquad k = \min\{\abs{I},n+1-\abs{I}\},
\end{equation}
where the combinatorial factor accounts for the permutations in $Sym_{n+1}$ which fix each $J$. As expected, $q'_I$ has the right dependence on cardinality only and we can now write $q'S\ge0$ using just $\tilde{S}_k$ variables. Declaring $q'_{I^\complement}=q'_{I}$ for convenience, one finds
\begin{equation}
    q'S = \sum_{\varnothing\ne I\subseteq[n]} q'_I S_I = \sum_{k=1}^{\ceil{n/2}} \sum_{I\in Q_k(n)} q'_I S_I = (n+1)! \sum_{k=1}^{\ceil{n/2}} \tilde{q}_k \tilde{S}_k,
\end{equation}
where in the last equality we have identified the desired \textit{symmetric coefficients} of our symmetrized inequality $\tilde{q}\tilde{S}\ge0$ to be
\begin{equation}
\label{eq:symi}
    \tilde{q}_k = \sum_{J\in Q_n(k)} q_J.
\end{equation}
Formally, one can understand this operation as the right inverse of the one in \eqref{eq:symn}. More explicitly, let $M$ and $N$ respectively be the rectangular matrices implementing \eqref{eq:symn} and \eqref{eq:symi} as linear maps, such that $\tilde{S}=MS$ and $\tilde{q} = N q$. Then one easily verifies that $N^T=M^+$, where $M^+ = M^T (M M^T)^{-1}$ is the canonical Moore–Penrose right inverse of $M$. For instance, for $n=3$ these matrices are
\begin{equation}
\label{eq:n3matrices}
    M=\begin{bmatrix}
        \frac{1}{4} & \frac{1}{4} & \frac{1}{4} & 0 & 0 & 0 & \frac{1}{4} \\
        0 & 0 & 0 & \frac{1}{3} & \frac{1}{3} & \frac{1}{3} & 0 
    \end{bmatrix}, \qquad
    N=\begin{bmatrix}
        1 & 1 & 1 & 0 & 0 & 0 & 1 \\
        0 & 0 & 0 & 1 & 1 & 1 & 0 
    \end{bmatrix},
\end{equation}
and one clearly has $MN^T=\mathds{1}_2$. Note in passing that these matrices can also be used to rephrase our symmetrization operations as projections of vectors $S\in\mathbb{R}^{2^n-1}$ onto a symmetric subspace within $\mathbb{R}^{2^n-1}$. In particular, such a projection is implemented by the matrix $N^TM$, which one can easily verify is idempotent as a projection should be. The result of applying $N^TM$ to any vector $S\in\mathbb{R}^{2^n-1}$ is to project it onto a subspace where, for all $I\subseteq[n]$, the coordinates attain values $S_{I}=\tilde{S}_{\abs{I}}$ if $\abs{I}\leq \ceil{n/2}$ and $S_{I}=\tilde{S}_{n+1-\abs{I}}$ otherwise. For the $n=3$ example in \eqref{eq:n3matrices}, the projection matrix just mentioned reads
\begin{equation}
\setlength\arraycolsep{4pt}
    N^TM=\begin{bmatrix}
        \frac{1}{4} & \frac{1}{4} & \frac{1}{4} & 0 & 0 & 0 & \frac{1}{4} \\
        \frac{1}{4} & \frac{1}{4} & \frac{1}{4} & 0 & 0 & 0 & \frac{1}{4} \\
        \frac{1}{4} & \frac{1}{4} & \frac{1}{4} & 0 & 0 & 0 & \frac{1}{4} \\
        0 & 0 & 0 & \frac{1}{3} & \frac{1}{3} & \frac{1}{3} & 0 \\
        0 & 0 & 0 & \frac{1}{3} & \frac{1}{3} & \frac{1}{3} & 0 \\
        0 & 0 & 0 & \frac{1}{3} & \frac{1}{3} & \frac{1}{3} & 0 \\
        \frac{1}{4} & \frac{1}{4} & \frac{1}{4} & 0 & 0 & 0 & \frac{1}{4}
    \end{bmatrix}.
\end{equation}

\section{The Symmetrized HEC}\label{sec:shec}

\begin{figure*}[]
	\begin{center}
		\includegraphics[width=7cm]{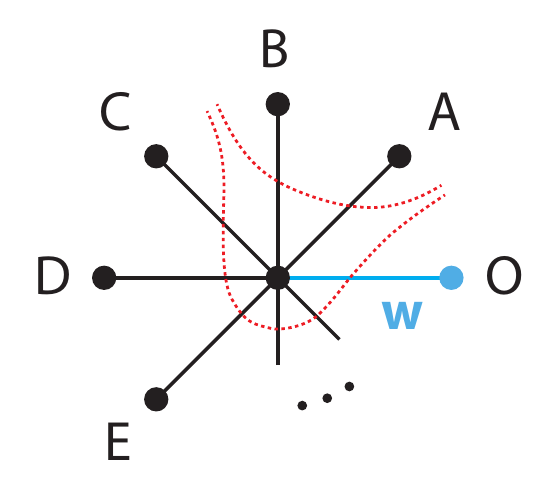}
	\end{center}
	\caption{General star graphs realizing the extreme rays of the SHEC. Black edges have unit weight, while the blue ``purification'' edge has positive integer weight. This is $w=1,3,\dots,n$ for $n$ odd, and $w=2,4,\dots,n$ for $n$ even. Dashed red lines represent the two possible min-cuts for subsystem $AB$. An analogous binary choice of including or excluding the central vertex exists for every subsystem.}
	\label{fig:star}
\end{figure*}

Full descriptions of the HEC in terms of extreme rays and facets are known for $n\leq5$ \cite{Bao:2015bfa,HernandezCuenca:2019wgh,Avis:2021xnz}. From this data, it is straightforward to compute the SHEC for $n\leq5$ by applying \eqref{eq:symn} to extreme rays and \eqref{eq:symi} to facets (see Appendix \ref{asec:exthec}). One can then easily check that the resulting descriptions of the SHEC are indeed dual to each other.

At this point, one already notices that the extreme rays of the SHEC for all $n\leq 5$ can be generated by a very simple family of graphs.\footnote{Intuitively, by ``generated'' here we mean that these graphs yield entropy vectors whose symmetrizations through \eqref{eq:symn} match the extreme rays of the SHEC.} In particular, they can all be captured by star graphs with $n+1$ edges, one for each boundary vertex, where the weight of all but one of them can be set to unity. The remaining edge, which by symmetry and without loss of generality can be picked to be the one associated to the purifier $n+1$, then carries some other integer weight $w\ge1$, as in Fig.~\ref{fig:star}. We observe that this family of star graphs turns out to yield precisely all extreme rays at each $n$ when the weight $w$ is chosen to be of the same parity as $n$ and obey $1\leq w \leq n$. These observations lead to the following conjecture:

\begin{nconj}[SHEC extreme rays]
\label{conj:shecrays}
    The extreme rays of the SHEC$_n$ can be generated by star graphs with $n$ weight-$1$ edges and another edge of positive weight $w=n,n-2,\dots$.
\end{nconj}

As a sanity check, notice that this proposal is consistent with the dimensionality of the SHEC, since the number of conjectured extreme rays $\ceil{n/2}$ precisely matches it. In turn, this suggests that the SHEC may be a simplicial cone for all $n$. To be more explicit, and building upon Conjecture \ref{conj:shecrays}, we now work out a closed-form expression for all conjectured extreme rays of the SHEC$_n$. Firstly, from the star graph in Fig.~\ref{fig:star}, every entropy $S_I$ is by definition given by
\begin{equation}
    S_I = \min \{ \abs{I},n-\abs{I}+w \} \;.
\end{equation}
Since $S_I$ only depends on the subsystem cardinality $k=\abs{I}$, using momentarily the notation $S_{I}=S_k$, it is then straightforward to apply \eqref{eq:symn} and obtain the coordinates of the symmetrized entropy vectors as
\begin{equation}
\label{eq:SHECrayktemp}
\begin{aligned}
    \tilde{S}_k 
    &= \binom{n+1}{k}^{-1} \left( \binom{n}{k} S_k + \binom{n}{n+1-k} S_{n+1-k} \right) \\
    &= \frac{1}{n+1} \left( (n+1-k)  \min \{ k,n-k+w \} + k  \min \{ n+1-k,w+k-1 \} \right) \;.
\end{aligned}
\end{equation}
The particular set of star graphs we are interested in for each $n$ can be conveniently parameterized by setting $w=n - 2(l-1)$ for $l=1,2,\dots,\ceil{n/2}$. This way, $l$ labels each of the $\ceil{l/2}$ conjectured extreme rays, whose entropies in \eqref{eq:SHECrayktemp} can be further simplified to
\begin{equation}
\label{eq:SHECrayk}
    \tilde{S}_k^{(l)} = \frac{2 k}{n+1} (n+1 - \max \{k,l\}).
\end{equation}
We can now use this result in order to obtain explicit results for the complete set of extreme rays of the SHEC for any number of parties $n$ according to Conjecture \ref{conj:shecrays}.

The $\ceil{n/2}$ entropy vectors of dimension $\ceil{n/2}$ we obtain for each $n$ this way can be arranged as columns of a square matrix. For instance, for $n=10$ (and with entropy vectors conveniently renormalized by a factor of $(n+1)/2$), this matrix can be written as
\begin{equation}
    M_{10}^{\text{SHEC}}=
    \renewcommand{\arraystretch}{1.1}
    \setlength\arraycolsep{3pt}
    \begin{bmatrix}
         10 & 9 & 8 & 7 & 6 \\
         18 & 18 & 16 & 14 & 12 \\
         24 & 24 & 24 & 21 & 18 \\
         28 & 28 & 28 & 28 & 24 \\
         30 & 30 & 30 & 30 & 30
    \end{bmatrix} \;.
\end{equation}
One can easily check that these matrices are non-singular for every $n$, thereby demonstrating that the conjectured SHEC indeed is a simplicial cone of full dimension in $\mathbb{R}^{\ceil{n/2}}$. As such, its dual facet description can be immediately obtained from the rows of the inverse of the matrices of extreme rays just described. For the $n=10$ example above, this leads to
\begin{equation}
\label{eq:eg10facets}
    (M_{10}^{\text{SHEC}})^{-1}=
    \begin{bmatrix}
         1 & -\frac{1}{2} & 0 & 0 & 0 \\
         -1 & ~~1 & -\frac{1}{3} & 0 & 0 \\
         0 & -\frac{1}{2} & ~~\frac{2}{3} & -\frac{1}{4} & 0 \\
         0 & 0 & -\frac{1}{3} & ~~\frac{1}{2} & -\frac{1}{5} \\
         0 & 0 & 0 & -\frac{1}{4} & ~~\frac{7}{30}
    \end{bmatrix},
\end{equation}
where from e.g. the first row one reads off the inequality
\begin{equation}
\label{eq:shecIneq1}
    2 \tilde{S}_{1} - \tilde{S}_{2} \geq 0 \;.
\end{equation}
This expression can be easily checked to come from a symmetrization of SA in \eqref{eq:sa} for singletons, and turns out to define a facet of the conjectured SHEC for all $n$.
Another inequality which shows up for every $n$ and involves symmetrized subsystems of just two different cardinalities is the following (cf. the last row in \eqref{eq:eg10facets}):
\begin{equation}
\label{eq:shecIneq2}
    -\left(1 -\frac{1}{\left\lfloor \frac{n+1}{2}\right\rfloor} \right)^{-1} \tilde{S}_{\ceil{n/2}-1} + \left( 1+\frac{1}{\left\lceil \frac{n+1}{2}\right\rceil } \right) \tilde{S}_{\ceil{n/2}} \geq 0 \;.
\end{equation}
Finally, the remaining set of inequalities completing the list of $\ceil{n/2}$ facets are captured by
\begin{equation}
\label{eq:shecIneq3}
    -l(l+1) \tilde{S}_{l-1} + 2 (l-1)(l+1) \tilde{S}_{l} - (l-1)l \tilde{S}_{l+1} \geq 0 \qquad\text{for}\quad l=2,\dots,\ceil{n/2}-1 \;.
\end{equation}
This last family can in fact be extended to $l=\ceil{n/2}$ in order to also reproduce \eqref{eq:shecIneq2} by declaring that $\tilde{S}_{\ceil{n/2}+1}=\tilde{S}_{\floor{n/2}} \;.$
To summarize, we have the following result:
\begin{ncor}[SHEC facets]
\label{cor:facets}
    If Conjecture \ref{conj:shecrays} holds, the facets of the SHEC$_n$ are defined by the set of inequalities given by \eqref{eq:shecIneq1}, \eqref{eq:shecIneq2}, and \eqref{eq:shecIneq3}.
\end{ncor}

Because all extreme rays in Conjecture \ref{conj:shecrays} are realizable by star graphs, it follows that the conjectured SHEC is contained in the true SHEC. Whether or not Conjecture \ref{conj:shecrays} is true thus depends on whether the conjectured facets of the SHEC are actually obtainable from symmetrizations of valid HEC inequalities. In other words, proving our conjecture requires finding and proving valid HEC inequalities for arbitrary $n$, which is a hard problem. Our explicit knowledge of the HEC$_n$ for small $n$ allows us to perform some checks on the conjectured SHEC. We will be able to provide a complete proof of our conjecture for up to $n=6$, but only a partial one for larger $n$.

As already mentioned, the conjectured SHEC facet \eqref{eq:shecIneq1} arises from singleton SA. Since this is a valid inequality of the HEC for all $n$, it follows that \eqref{eq:shecIneq1} is a valid facet of the SHEC for all $n$. Consider now \eqref{eq:shecIneq3} for $l=2$, which yields
\begin{equation}
\label{eq:mmisym}
    -3\tilde{S}_1 + 3\tilde{S}_2 - \tilde{S}_3 \ge 0 \;.
\end{equation}
It is a simple exercise to show that this follows from the MMI inequality in \eqref{eq:mmi} for singletons when symmetrized for $n\ge 5$.\footnote{For $n=3$ and $n=4$, as shown in Appendix \ref{asec:exthec}, \eqref{eq:mmisym} respectively collapses down to $-4\tilde{S}_1 + 3\tilde{S}_2 \ge 0$ and $-3\tilde{S}_1 + 2\tilde{S}_2 \ge 0$, which take the form of \eqref{eq:shecIneq2} (see comment below \eqref{eq:shecIneq3}).}
Again, since MMI is a well-known HEC inequality which holds for all $n$, it follows that \eqref{eq:mmisym} is a valid facet of the SHEC for all $n$. Further, \eqref{eq:shecIneq3} for $l=3$ gives
\begin{equation}
\label{eq:5sym}
    -6\tilde{S}_2 + 8\tilde{S}_3 - 3\tilde{S}_4 \ge 0 \;.
\end{equation}
This inequality can be seen to arise from any one of the last three $5$-party inequalities in Appendix \ref{asec:exthec} when symmetrized for $n\ge7$.\footnote{As was the case for MMI in the footnote above, if symmetrized for $n=5$ or $n=6$, these instead give rise to the special case of \eqref{eq:shecIneq2} corresponding to \eqref{eq:shecIneq3} applied to $l=\ceil{n/2}$. For $n=5$, as shown in Appendix \ref{asec:exthec}, these symmetrize down to $-9\tilde{S}_2+8\tilde{S}_3\ge0$, while for $n=6$ they give $-6\tilde{S}_2+5\tilde{S}_3\ge0$.}

These results suffice to prove validity of our proposal for the extremal elements of the SHEC$_n$ for up to $n=6$. Unfortunately, our incomplete knowledge of the HEC$_n$ for larger $n$ is insufficient to do so for $n\geq 7$, for which a different approach that bypasses full knowledge of the HEC will be needed.

\section{The Symmetrized QEC}\label{sec:sqec}

The QEC is poorly understood and only known exactly for up to $n=3$. At this number of parties, its facets belong to the two orbits associated to the following inequalities:
\begin{equation}
    \begin{aligned}
        S_{1}+S_{2} & \geq S_{12} \;, \\
        S_{12}+S_{23} & \geq S_{2}+S_{123} \;.        
    \end{aligned}
\end{equation}
The first one can be recognized as the SA inequality in \eqref{eq:sa}, while the last one corresponds to SSA as in \eqref{eq:ssa}. Going to higher $n$, different instances of the same type of inequality associated to inequivalent choices of subsystems can define algebraically independent orbits of inequalities. This way, one can apply SA and SSA to all possible subsystems in order to construct a rich polyhedral cone bounded by multiple orbits of these two types of inequalities. Because both are universal quantum inequalities, the resulting cone defines an outer approximation to the QEC. A neat study of precisely which of these orbits of SA and SSA constitute facets of the resulting cone, and which are redundant, was performed by Nicholas Pippenger in \cite{pippenger2003inequalities}.

Additionally, for arbitrary number of parties $n$, Pippenger studied the cone that results under $Sym_n$ symmetrizations. More specifically, rather than symmetrizing over all $I\subseteq[n+1]$ for each $\abs{I}=k\leq\ceil{n/2}$, Pippenger considered symmetrizations only over $I\subseteq[n]$ for each $\abs{I}=k\leq n$, thus not including the purification symmetry. Even for this milder $Sym_n$ symmetrizations, he was able to prove that every extreme ray of the resulting cone is realizable by some quantum state.\footnote{There is a minor typo associated to Theorem $5.5$ of \cite{pippenger2003inequalities}: in equation $(5.20)$ therein, the condition $n-a+1\le b\le n$ should instead read $n-a\le b\le n$.} This way, he established that the cone of SA and SSA instances collapses down to precisely match the QEC under respective $Sym_n$ symmetrizations. As we did for the HEC, here we will instead consider $Sym_{n+1}$ symmetrizations, i.e., we
further symmetrize over the purifier. 
This symmetrization of the QEC defines the SQEC, which by \cite{pippenger2003inequalities} will of course continue to match the cone of SA and SSA instances accordingly symmetrized under $Sym_{n+1}$. In other words, using the results of \cite{pippenger2003inequalities}, we will be able to present complete descriptions of the SQEC for all $n$.

\ignore{
Consider here the cone defined by SA and SSA. For more parties, facets of this object are just defined by these inequalities after permutation of the parties.

Pippenger gives the list of inequalities
\begin{equation}
\label{eq:sassaPip}
    \begin{cases}
    - S_{i-1} + 2 S_{i} - S_{i+1} \geq 0 &\qquad \forall\;  1\leq i\leq n-1 \\
    - S_{i-1} - S_{n-i} + S_i + S_{n-i+1} \geq 0 &\qquad \forall\;  1\leq i\leq n-1 \\
    \end{cases}
\end{equation}
with $S_0=0$. Note that the second family of inequalities is unchanged if we replace $i$ by $n-i+1$, meaning that there are only $\ceil{n/2}$ distinct inequalities.
This gives in total $(n-1)+\ceil{n/2}$ inequalities. From these, the extreme rays can be obtained using dedicated software.

The extreme rays of the symmetric quantum cone are know explicitly for any n (the symmetric quantum cone is actually the symmetric SSA cone, see Pippenger). These take the form of Eq.(5.5) in his paper, with $a,b$ satisfying Eq.(5.20), as stated in Theorem 5.5.

Following the notation in Pippenger, consider states $\rho^{a,b}$ with
\begin{equation}
    1 \leq a \leq n \qquad\qquad \max(a,n-a) \leq b \leq n \;.
\end{equation}
The associated entropy is
\begin{equation}
    S(\rho^{a,b}) =  \begin{cases}
    i m & \text{if $0 \leq i \leq a$} \\
    a m & \text{if $a+1 \leq i \leq b$} \\
    (a+b-i) m & \text{if $b+1 \leq i \leq n$} 
  \end{cases}
\end{equation}
where $m=\ceil{\log_2(2N+1)}$

}

Like the SHEC, the SQEC turns out to be simplicial. For each $n$, the facets of the SQEC are given by the following set of $\ceil{n/2}$ inequalities:
\begin{equation}
\label{eq:sassafull}
    - \tilde{S}_{l-1} + 2 \tilde{S}_{l} - \tilde{S}_{l+1} \geq 0 \qquad \text{for} \quad  1\leq l\leq \ceil{n/2} \;,
\end{equation}
where we have declared $\tilde{S}_0=0$ and $\tilde{S}_{\ceil{n/2}+1} = \tilde{S}_{\ceil{n/2}}$. For $l=1$, \eqref{eq:sassafull} yields \eqref{eq:shecIneq1}, which recall comes from the SA facet that the QEC shares with the HEC. The rest come from symmetrizations of SSA inequalities as in \eqref{eq:ssa} for subsystems $I,J\subseteq[n]$ obeying $\abs{I \cap J} = 2,3,\dots,\ceil{n/2}$.

One can easily see that the SQEC is a simplicial cone. Writing out its facet-defining vectors as rows in a matrix, the extreme rays of the SQEC can thus be obtained as the column vectors of the inverse matrix. For instance, for $n=10$, the matrix of row inequalities reads
\begin{equation}
    M_{10}^{\text{SQEC}} = 
    \begin{bmatrix}
         2 & -1 & 0 & 0 & 0 \\
         -1 & 2 & -1 & 0 & 0 \\
         0 & -1 & 2 & -1 & 0 \\
         0 & 0 & -1 & 2 & -1 \\
         0 & 0 & 0 & -1 & 1
    \end{bmatrix} \;,
\end{equation}
and its inverse of column extreme rays yields
\begin{equation}\label{eq:SQECM10}
    (M_{10}^{\text{SQEC}})^{-1} = 
    \setlength\arraycolsep{6pt}
    \begin{bmatrix}
         1 & 1 & 1 & 1 & 1 \\
         1 & 2 & 2 & 2 & 2 \\
         1 & 2 & 3 & 3 & 3 \\
         1 & 2 & 3 & 4 & 4 \\
         1 & 2 & 3 & 4 & 5 \\
    \end{bmatrix} \;.
\end{equation}
For general $n$, letting $l=1,2,\dots,\ceil{n/2}$ enumerate each of the resulting vectors, the $k^{\text{th}}$ component of the $l^{\text{th}}$ extreme ray can be easily seen to be given by
\begin{equation}
\label{eq:SQECrays}
    \tilde{S}^{(l)}_k = \min\{k,l\} \;.
\end{equation}

\begin{figure*}
	\begin{center}
		\includegraphics[width=\textwidth]{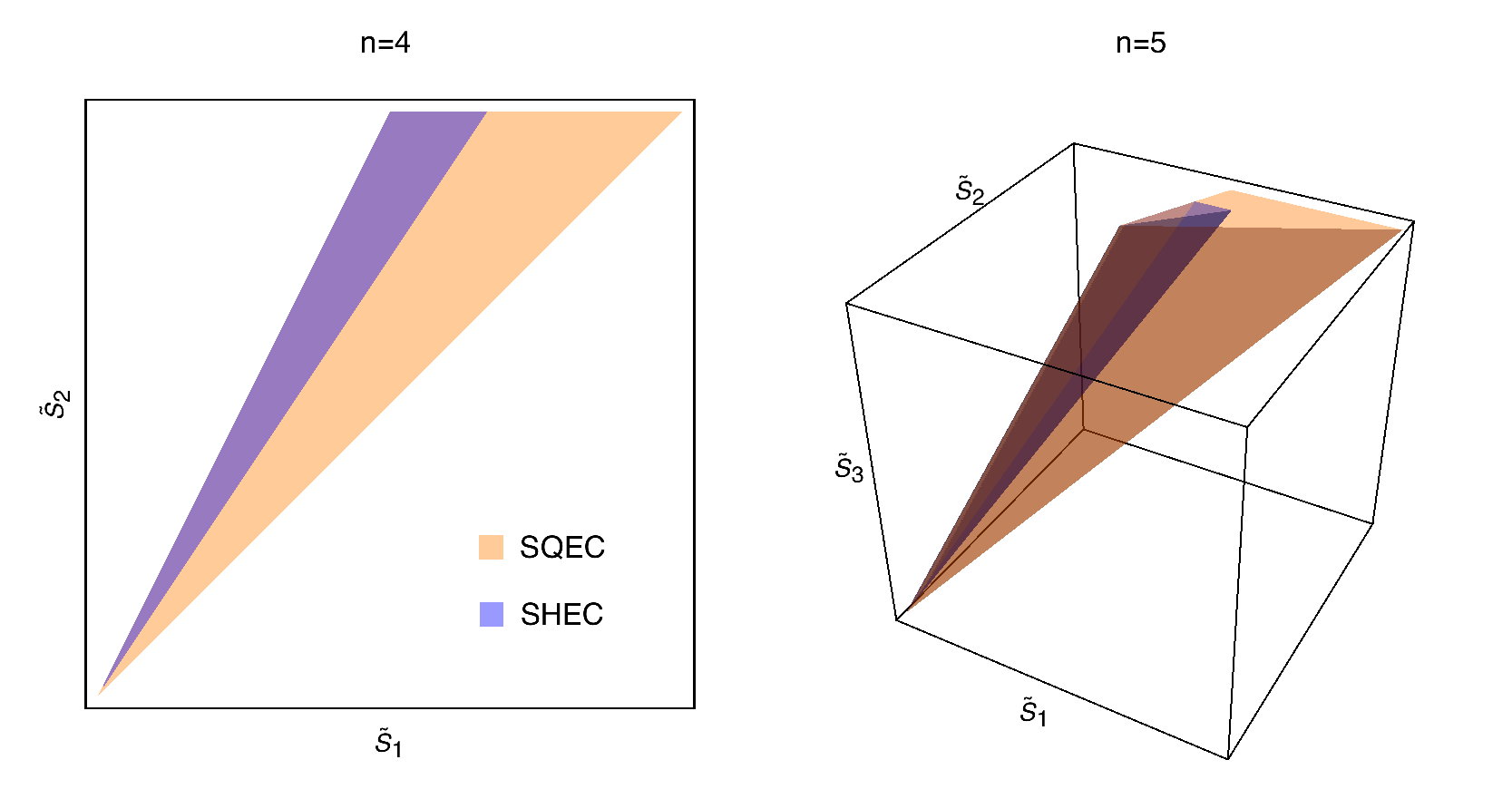}
	\end{center}
	\caption{The SQEC and SHEC for $n=4,5$.}
	\label{fig:cones}
\end{figure*}

\begin{figure*}
	\begin{center}
	    \hfill
		\includegraphics[width=.3\textwidth]{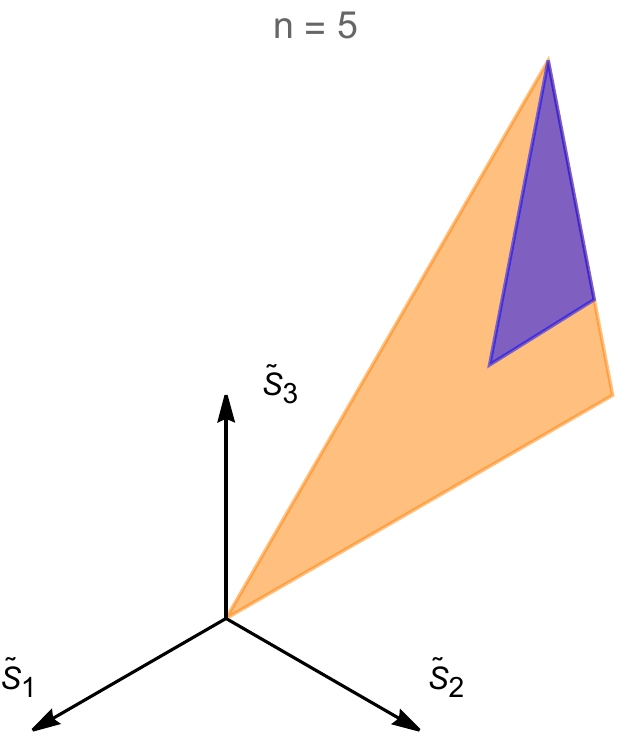}
		\hfill
		\includegraphics[width=.3\textwidth]{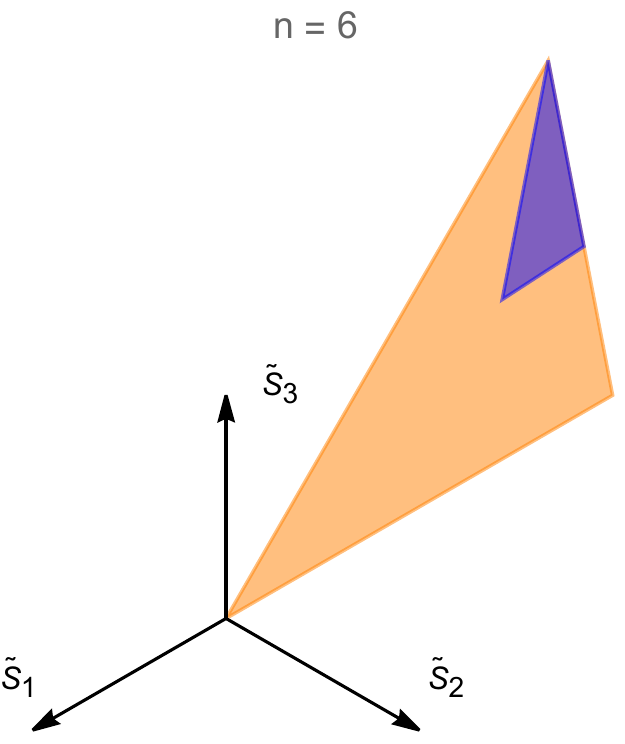}
		\hfill\\\vspace{15pt}
		\hfill
		\includegraphics[width=.3\textwidth]{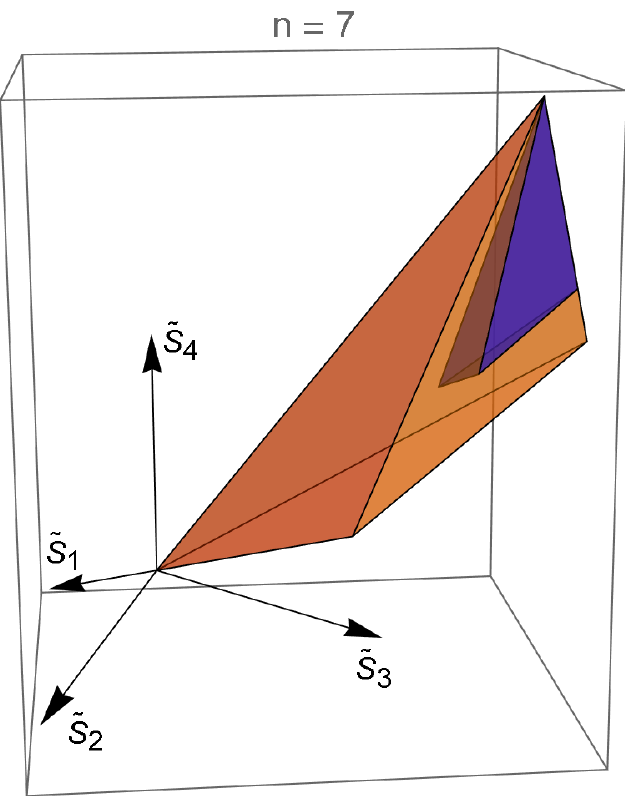}
		\hfill
		\includegraphics[width=.3\textwidth]{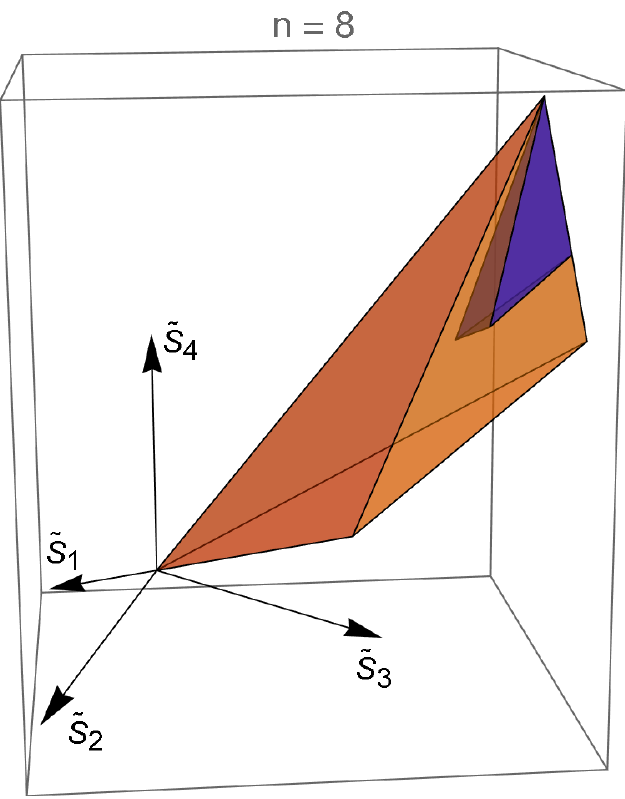}
		\hfill\\
	\end{center}
	\caption{Cross-sections of the SQEC and SHEC through the hyperplane defined by $\sum_{k=1}^{\ceil{n/2}} \tilde{S}_k = 1$ for $n=5,6,7,8$. The cross-section of the positive orthant is not shown because it is too large. Instead, for orientation we show arrows pointing in the direction in which the intersections with the canonical basis vectors occur -- these would define the extreme points of the positive orthant in the cross-section.}
	\label{fig:conescross}
\end{figure*}

In Figure~\ref{fig:cones}, we illustrate the SQEC and SHEC full-dimensionally for $n=4,5$. In Figure~\ref{fig:conescross}, we illustrate the SQEC and SHEC via codimension-$1$ radial cross-sections for $n=5,6,7,8$. A question naturally arising is now how these two cones relate to each other. In particular, we will be interested in comparing their ``size'', and the scaling with $n$.

\section{Volumes of Symmetrized Entropy Cones}\label{sec:vols}

The volume of a pointed polyhedral cone is infinite. However, if the cone lies within a small solid angle inside the positive orthant, then a useful notion of how large it is can be obtained as the volume of the polytope it forms up to some choice of cross-section. When this cone is simplicial, as is the case for both the SQEC and the SHEC, this polytope is a simplex whose vertices are the origin and the intersection of its extreme rays with the cross-section. The volume of such a simplex can be easily computed using the determinant of the matrix of extreme rays $\hat{M}$ suitably normalized to match the vertices on the cross-section. Since the determinant actually computes the volume of the parallelotope formed by the vectors involved, one has to divide it by a combinatorial factor of $d!$ to obtain the volume of the corresponding $d$-dimensional simplex. Denoting a general matrix of normalized extreme rays of a $d$-dimensional simplicial cone by $\hat{M}$, we will thus define its volume by
\begin{equation}
\label{eq:stravol}
    \vol\left( \cone\,\{\hat{M}\} \right) \equiv \frac{1}{d!} |\det \hat{M}| \;.
\end{equation}

When considering the SQEC and the SHEC, our starting point will be a matrix $M$ of unnormalized, integral extreme rays.
A natural choice of cross-section is the hyperplane defined by $\sum_{k=1}^{\ceil{n/2}}\tilde{S}_k=1$, as in Fig.~\ref{fig:conescross}. The vertices on the cross-section can then be easily obtained by normalizing the extreme rays of the desired cone linearly to unity, i.e., using the taxicab norm. Explicit numerical results for small $n$ are given in Appendix \ref{sec:volSmallN}. More generally, we are interested in the scaling of volumes as $n\rightarrow\infty$. Our strategy for obtaining this will be as follows:
\begin{enumerate}
    \item\label{s1} Derive a simple general expression for the determinant of $M$.
    \item\label{s2} Derive an expression for the product $\Pi(M)$ of the norms of all extreme rays in $M$.
    \item\label{s3} Obtain the desired (normalized) determinant as $|\det\hat{M}| = \frac{\abs{\det M}}{\Pi(M)}$.
\end{enumerate}

For the SQEC$_n$, the extreme rays are given component-wise by \eqref{eq:SQECrays}, which we will take to define our unnormalized matrix of extreme rays $M^{\text{SQEC}}_n$. The determinant of the matrix formed by these vectors is always unity for all $n$, i.e., $\det M^{\text{SQEC}}_n =1$, which takes care of Step \ref{s1}. Letting $d_n=\ceil{n/2}$ for convenience, the normalization factor $\Pi(M^{\text{SQEC}}_n)$ resulting from Step \ref{s2} can also be obtained exactly and reads
\begin{equation}
\label{eq:sqcnorm}
    \Pi(M^{\text{SQEC}}_n) = \frac{\left(2 d_n \right)!}{2^{d_n }}.
\end{equation}
We can thus conclude that the volume of the SQEC for $n$ parties is exactly given by
\begin{equation}
\label{eq:volSQEC}
    \vol(\text{SQEC}_n) = \frac{1}{d_n!} \frac{2^{d_n }}{\left(2 d_n \right)!},
\end{equation}
Since we will be ultimately interested only in the large-$n$ scaling of this volume, we may consider keeping track of just the leading order behavior of \eqref{eq:volSQEC} as $n\to\infty$. Undergoing numerically coarser approximations at large $n$, and taking $n$ to be even (such that $d_n = n/2$), one obtains
\begin{equation}
\label{eq:volSQECapp}
    \vol(\text{SQEC}_n) \sim \frac{1}{(n/2)!}\frac{e^{\frac{13}{10} n}}{n^{\frac{1}{2}+n}} \sim \frac{e^{\frac{11}{5}n}}{n^{1+ \frac{3}{2}n}}.
\end{equation}

Although slightly more complicated, the same approach can be applied to the SHEC. We will use \eqref{eq:SHECrayk} with an $(n+1)/2$ prefactor for the unnormalized extreme rays in $M^{\text{SHEC}}_n$.
Remarkably, the determinants of the resulting matrix are easily computable and read
\begin{equation}
    \det M^{\text{SHEC}}_n =
    \left\lceil{\frac{n+1}{2}}\right\rceil! \times
    \begin{cases}
    1&\quad \text{if $n$ is even,}\\
    \frac{n+1}{2}&\quad \text{if $n$ is odd.}
    \end{cases}
\end{equation}
The exact result for $\Pi(M^{\text{SHEC}}_n)$ can be written
\begin{equation}
\label{eq:shecnorm}
    \Pi(M^{\text{SHEC}}_n) = \left(-\frac{1}{6}\right)^{d_n} \prod_{k=1}^3 (1-x_k)_{d_n} \;,
\end{equation}
where the subscripted brackets denote the Pochhammer symbol $(a)_k = \frac{\Gamma(a+k)}{\Gamma(a)}$, and $x_k$ stands for the $k^{\text{th}}$ root of the following $3^{\text{rd}}$-order polynomial:
\begin{equation}
\label{eq:poly5}
    \prod_{k=1}^3 (x-x_k) = x^3-x-d_n (d_n+1) (3 n-2 (d_n-1)).
\end{equation}
Albeit exact, \eqref{eq:shecnorm} still relies on an implicit expression for the $x_k$ roots. We could of course use the general formulae for the roots of a $3^{\text{rd}}$-order polynomial, but this would not be very illuminating. Instead, we will take advantage of large-$n$ approximations in order to get a more explicit result for this. A key observation is to notice that the roots of \eqref{eq:poly5} scale linearly with $n$. Hence, to simplify this polynomial non-trivially at large $n$, we will first have to factor out this uninteresting $n$-dependence. We can do so by letting $x = d_n \hat{x}$ and introducing renormalized roots $x_k=d_n\hat{x}_k$. Plugging these into \eqref{eq:poly5}, at leading order in $n$ we obtain
\begin{equation}
\label{eq:simpoly5}
    \prod_{k=1}^3 (\hat{x}-\hat{x}_k) \approx \hat{x}^3-4 \;.
\end{equation}
The renormalized roots no longer depend on $n$ and are simply the cubic roots of $4$. In terms of these, we can re-express \eqref{eq:shecnorm} as
\begin{equation}
\label{eq:ab}
    \Pi(M^{\text{SHEC}}_n) \approx \left(-\frac{1}{6}\right)^{d_n} \prod_{k=1}^3 (1- \hat{x}_k d_n)_{d_n}\;,
\end{equation}
and expand the Pochhammer symbols at large $n$. For each root, this expansion yields
\begin{equation}
    (1-d_n \hat{x}_k)_{d_n} \approx \left(1-\frac{1}{\hat{x}_k}\right)^{\frac{1}{2}+(1-\hat{x}_k) d_n} \left(-\frac{\hat{x}_k d_n}{e}\right)^{d_n}.
\end{equation}
Taking the profuct over all roots in \eqref{eq:ab} and making some numerical approximations, we arrive at
\begin{equation}
    \Pi(M^{\text{SHEC}}_n) \approx \frac{\sqrt{3}}{2} \left(\frac{3}{5}\right)^{d_n} d_n^{3 d_n} \sim e^{-\frac{13}{10} n} n^{\frac{3}{2} n}.
\end{equation}
Having completed Step \ref{s2}, we finally obtain our desired expression for the volume of the SHEC at large $n$:
\begin{equation}
\label{eq:volSHECapp}
    \vol(\text{SHEC}_n) \sim \frac{1}{(n/2)!} \frac{e^{\frac{9}{2} n}}{n^{-1+n}} \sim \frac{e^{\frac{13}{10} n}}{n^{-\frac{1}{2}+\frac{3}{2}n}}\;.
\end{equation}

We are finally ready to compare the asymptotic volumes of the SQEC and the SHEC. Using the expressions from \eqref{eq:volSQECapp} and \eqref{eq:volSHECapp}, we obtain
\begin{equation}\label{eq:volRatio}
    \frac{\vol(\text{SHEC}_n)}{\vol(\text{SQEC}_n)} \sim n^{\frac{3}{2}} e^{-\frac{9}{10} n}.
\end{equation}
We thus conclude that, in terms of volumes in entropy space, the SHEC$_n$ constitutes an exponentially small fraction of the SQEC$_n$. This is particularly interesting considering that holographic states are in fact expected to be typical when randomly sampling some Hilbert space of quantum states with respect to an invariant measure on it.\footnote{We thank Bartek Czech for comments on this points.}
We can also compare the volumes of these two cones to the volume of the unit simplex that the positive orthant defines. Using $n/2$ as the dimensionality at large $n$, the volume of this object is simply given by $1/(n/2)!$. Comparing this to \eqref{eq:volSQEC}, we see that the SQEC$_n$ itself also occupies a very small fraction of the positive orthant which decreases faster than exponential as $n$ increases.

\section{Conclusion and Outlook}\label{sec:end}

We motivated and presented a projection of the HEC via a natural symmetrization which results in a much simpler (but still non-trivial) entropy cone: the SHEC. By inspecting its extremal elements, we were able to conjecture a full characterization of this cone for an arbitrary number of parties $n$ in terms of both full sets of extreme rays and facet inequalities.
Our proof of this proposal for the SHEC for small $n\le6$ was based on the known complete descriptions of the HEC for $n\leq 5$.
In the future, however, since the SHEC is by construction a coarser and much simpler version of the HEC, one would hope to make progress in proving the general form of the SHEC without necessarily knowing such a description for the HEC. Indeed, already at the conjectural level, here we have provided a plausible general form of the SHEC for arbitrary $n$, while such an achievement for the HEC seems unattainable as of now.

For instance, the partial description of the HEC$_6$ known to date \cite{n6wip} already exhibits an unprecedented level of complexity compared to $n\le5$. While the HEC$_5$ consists of just $8$ distinct orbits of facets and $19$ orbits of extreme rays, the current description of the HEC$_6$ already involves at least $182$ orbits of facets and at least $4122$ orbits of extreme rays \cite{n6wip}.\footnote{Data to date suggests that, in fact, these numbers are still highly underestimating the complexity of a complete description of the HEC$_6$.} According to Conjecture \ref{conj:shecrays} and Corollary \ref{cor:facets}, all this extremely rich structure that arises for $n\ge6$ turns out to be symmetrized away at the level of the SHEC, which nonetheless still preserves some non-trivial properties of the HEC in the form of $\ceil{n/2}$ facets and extreme rays at each $n$.

Symmetrizations provide an organizing principle for the complicated structure of the HEC. The usefulness of this organization is already evidenced by $n=5$ e.g. at the level of facets, where we observe that $1)$ not all facets of the HEC remain facets of the SHEC, $2)$ those which remain facets do so in non-trivial families of multiple facets, and $3)$ it takes genuinely new facets of the HEC at each $n$ to generate higher-$n$ facets of the SHEC. Observation $1)$ is consistent with our expectations for the SHEC: for larger $n$ there arise combinatorially large numbers of new orbits of facets for the HEC, but the simplicial SHEC should only involve $\ceil{n/2}$ as per our conjecture. Observation $2)$ implies that there are distinguished families of facets of the HEC which remain ``extremal'' upon symmetrization. Observation $3)$ textitasizes the fact that the SHEC retains non-trivial information about the HEC for all $n$, thus stressing its usefulness as a prerequisite to understand the HEC. %
To elaborate further on this, let us textitasize that none of the $182$ facet orbits that we currently known of the HEC$_6$ happens to prove the form of the SHEC for $n=7$ -- namely, none of them symmetrizes down to the $l=4$ inequality coming from \eqref{eq:shecIneq3}. This suggests that genuinely $n=7$ information about the HEC may be needed to characterize the SHEC$_7$, and similarly for any larger $n$. It is also interesting to note that even though the $l=3$ inequality coming from \eqref{eq:shecIneq3} can be realized by HEC$_5$ inequalities, there also appear new HEC$_6$ inequalities falling into the same family. For instance,
\begin{equation}
\begin{aligned}
    -S_{12}-S_{13}-S_{14}-&S_{25}-S_{26}-S_{35}+S_{123}+S_{124}+S_{125}+S_{126}\\+&S_{134}+S_{135}+S_{235}+S_{256}-S_{1234}-S_{1235}-S_{1256} \ge 0 \;,
\end{aligned}
\end{equation}
is a genuinely HEC$_6$ inequality (i.e., not obtainable as a lift from lower-$n$ inequalities), whose symmetrization for $n\ge7$ matches precisely \eqref{eq:5sym} as well.

More generally, the symmetrizations studied here have proven to be useful in distilling non-trivial structure at arbitrary $n$ for both the quantum and holographic entropy cones. Other interesting classes of quantum states and constructs have been studied in the past, which suffer from the same rapidly increasing complexity as $n$ increases. It would be interesting to explore if, upon symmetrizations, one can gain some further knowledge about the general-$n$ structure of the cone of stabilizer states \cite{Linden:2013kal}, the cone of linear rank inequalities \cite{Dougherty2009}, the cone of hypergraph entropies \cite{Bao:2020zgx,Walter:2020zvt,Bao:2020mqq}, the cone of topological links \cite{Bao:2021gzu}, or even the HEC under quantum corrections from bulk matter fields \cite{Akers:2021lms}.

\acknowledgments{It is a pleasure to thank David Avis, Max Rota and Jordi Tura for useful discussions. 
MF was supported by the Swiss National Science Foundation and by The Branco Weiss Fellowship -- Society in Science, administered by the ETH Z\"{u}rich.
SHC was supported by NSF grant PHY-1801805 and UCSB during the early stages of this project, and is currently supported by NSF grant PHY-2107939 and a Len DeBenedictis Graduate Fellowship.}

\vspace{10mm}
\noindent\textbf{Note --} During the completion of this work we became aware of \cite{Czech:2021rxe}, which applies the same projection to the HEC and obtains results consistent with ours.


\addcontentsline{toc}{section}{References}
\bibliographystyle{JHEP}
\bibliography{references.bib}

\clearpage
\newpage

\appendix

\section{Extremal structure of the SHEC for $n\leq 5$}\label{asec:exthec}

The HEC has been completely characterized for $n\leq 5$ \cite{Bao:2015bfa,HernandezCuenca:2019wgh,Avis:2021xnz}, while for larger $n$ only partial results are known due to its rapidly increasing complexity \cite{n6wip}. We list here all the extreme rays and facets of the HEC for $n\leq 5$, and compute their symetrization according to the prescription presented in Section~\ref{sec:symm}.

\subsection*{Extreme Rays}\label{supp:raysUptoN5}

In the tables below, the column ``extremal?'' refers to whether the symmetrized ray is extremal for the associated SHEC, while the column ``$w$'' indicates the weight of the purification edge in the star graph realizing the corresponding ray (see Fig.~\ref{fig:star}).

\begin{table}[h!]
\centering
$n=2$\\
\vspace{2mm}
\begin{tabular}{|r|c|c|c|c|}
\hline
$\#$ & HEC & SHEC & extremal? & $w$\\ \hline
1 & (11; 0) & (1) & yes & any\\ \hline
\end{tabular}
\end{table}

\begin{table}[h!]
\centering
$n=3$\\
\vspace{2mm}
\begin{tabular}{|r|c|c|c|c|}
\hline
$\#$ & HEC & SHEC & extremal? & $w$\\ \hline
1 & (110; 011; 0) & (3 4) & yes & 3\\ \hline
2 & (111; 222; 1) & (1 2) & yes & 1\\ \hline
\end{tabular}
\end{table}

\begin{table}[h!]
\centering
$n=4$\\
\vspace{2mm}
\begin{tabular}{|r|c|c|c|c|}
\hline
$\#$ & HEC & SHEC & extremal? & $w$\\ \hline
1 & (1100; 011110; 0011; 0) & (2 3) & yes & 4\\ \hline
2 & (1110; 221211; 1222; 1) & (1 2) & yes & 2\\ \hline
3 & (1111; 222222; 3333; 2) & (1 2) & yes & 2\\ \hline
\end{tabular}
\end{table}

\begin{table}[h!]
\centering
$n=5$\\
\vspace{2mm}
\begin{tabular}{|r|c|c|c|c|}
\hline
$\#$ & HEC & SHEC & extremal? & $w$\\ \hline
1 & (11000; 0111111000; 0001111110; 00011; 0) & (5 8 9) & yes & 5\\ \hline

2 & (11100; 2211211110; 1222212211; 11222; 1) & (5 10 12) & & \\ \hline
3 & (11110; 2221221211; 3323223222; 23333; 2) & (5 10 12) & & \\ \hline

4 & (11111; 2222222222; 2233333323; 22222; 1) & (10 20 27) & & \\ \hline
5 & (11111; 2222222222; 3332233323; 22222; 1) & (10 20 27) & & \\ \hline
6 & (11112; 2223223233; 3343443444; 43333; 2) & (10 20 27) & & \\ \hline

7 &  (11111; 2222222222; 2222333333; 22222; 1) & (5 10 13) & & \\ \hline
8 &  (11111; 2222222222; 2233223333; 22222; 1) & (5 10 13) & & \\ \hline
9 &  (33333; 6666666666; 5777799999; 66666; 3) & (5 10 13) & & \\ \hline
10 & (33333; 6666666666; 5777979999; 66666; 3) & (5 10 13) & & \\ \hline
11 & (33333; 6666666666; 5777999799; 66666; 3) & (5 10 13) & & \\ \hline

12 & (11111; 2222222222; 1333333333; 22222; 1) & (5 10 14) & & \\ \hline

13 & (11111; 2222222222; 3333333333; 22222; 1) & (1 2 3) & yes & 1 \\ \hline

14 & (11111; 2222222222; 3333333333; 44444; 3) & (4 8 9) & yes & 3 \\ \hline

15 & (11112; 2223223233; 3323443444; 43333; 2) & (20 40 51) & & \\ \hline

16 & (11122; 2233233334; 3444454455; 55444; 3) & (25 50 63) & & \\ \hline
17 & (11222; 2333333444; 4445355534; 44433; 2) & (25 50 63) & & \\ \hline

18 & (22223; 4445445455; 4656576777; 65555; 3) & (7 14 18) & & \\ \hline
19 & (22223; 4445445455; 4656756777; 65555; 3) & (7 14 18) & & \\ \hline

\end{tabular}
\end{table}

\clearpage
\newpage

\subsection*{Facets} \label{supp:facetsUptoN5}

Below, the table entry ``facet?'' refers to whether the symmetrized inequality is extremal for the associated SHEC.

\begin{table}[h!]
\centering
$n=2$\\
\vspace{2mm}
\begin{tabular}{|r|c|c|c|}
\hline
$\#$ & HEC & SHEC & facet? \\ \hline
1 & $\S{A} + \S{B} - \S{AB} \geq 0$ & $\tilde{S}_1 \geq 0$ & yes \\ \hline
\end{tabular}
\end{table}

\begin{table}[h!]
\centering
$n=3$\\
\vspace{2mm}
\begin{tabular}{|r|c|c|c|}
\hline
$\#$ & HEC & SHEC & facet? \\ \hline
1 & $\S{A} + \S{B} - \S{AB} \geq 0$ & $2 \tilde{S}_1 - \tilde{S}_2 \geq 0$ & yes \\ \hline
2 & $\S{AB} + \S{AC} + \S{BC} - \S{A} - \S{B} - \S{C} - \S{ABC} \geq 0$ & $-4 \tilde{S}_1 + 3\tilde{S}_2 \geq 0$ & yes \\ \hline
\end{tabular}
\end{table}

\begin{table}[h!]
\centering
$n=4$\\
\vspace{2mm}
\begin{tabular}{|r|c|c|c|}
\hline
$\#$ & HEC & SHEC & facet? \\ \hline
1 & $\S{A} + \S{B} - \S{AB} \geq 0$ & $2 \tilde{S}_1 - \tilde{S}_2 \geq 0$ & yes \\ \hline
2 & $\S{AB}+\S{AC}+\S{BC} -\S{A}-\S{B}-\S{C}-\S{ABC} \geq 0$ & $-3 \tilde{S}_1 + 2 \tilde{S}_2 \geq 0$ & yes \\ \hline
\end{tabular}
\end{table}

\begin{table}[h!]
\centering
$n=5$\\
\vspace{2mm}
\begin{tabular}{|c|m{.6\textwidth}|c|c|}
\hline
$\#$ & HEC & SHEC & facet? \\ \hline
1 & $\S{A} + \S{B} - \S{AB} \geq 0$ & $2 \tilde{S}_1 - \tilde{S}_2 \geq 0$ & yes \\ \hline
2 & $\S{AB}+\S{AC}+\S{BC} -\S{A}-\S{B}-\S{C}-\S{ABC} \geq 0$ & $-3 \tilde{S}_1 + 3 \tilde{S}_2 - \tilde{S}_3 \geq 0$ & yes \\ \hline
3 & $\S{AB}+\S{ACD}+\S{BCD} -\S{A}-\S{B}-\S{CD}-\S{ABCD} \geq 0$ & $-2 \tilde{S}_1 - \tilde{S}_2 + 2\tilde{S}_3 \geq 0$ &  \\ \hline
4 & $\S{AD}+\S{BC}+\S{ABE}+\S{ACE}+\S{ADE}+\S{BDE}+\S{CDE} -\S{A}-\S{B}-\S{C}-\S{D}-\S{AE}-\S{DE}-\S{BCE}-\S{ABDE}-\S{ACDE} \geq 0$ & $-2 \tilde{S}_1 - \tilde{S}_2 + 2\tilde{S}_3 \geq 0$ &  \\ \hline
5 & $\S{ABC}+\S{BCD}+\S{CDE}+\S{ADE}+\S{ABE} -\S{AB}-\S{BC}-\S{CD}-\S{DE}-\S{AE}-\S{ABCDE} \geq 0$ & $- \tilde{S}_1 -5 \tilde{S}_2 + 5\tilde{S}_3 \geq 0$ &  \\ \hline
6 & $2 \S{ABC}+\S{ABD}+\S{ABE}+\S{ACD}+\S{ADE}+\S{BCE}+\S{BDE} -\S{AB}-\S{AC}-\S{AD}-\S{BC}-\S{BE}-\S{DE}-\S{ABCD}-\S{ABCE}-\S{ABDE} \geq 0$ & $-9 \tilde{S}_2 + 8\tilde{S}_3 \geq 0$ & yes \\ \hline
7 & $\S{ABC}+\S{ABD}+\S{ABE}+\S{ACD}+\S{ACE}+\S{ADE}+\S{BCE}+\S{BDE}+\S{CDE}-\S{AB}-\S{AC}-\S{AD}-\S{BE}-\S{CE}-\S{DE}-\S{BCD}-\S{ABCE}-\S{ABDE}-\S{ACDE} \geq 0$ & $-9 \tilde{S}_2 + 8\tilde{S}_3 \geq 0$ & yes \\ \hline
8 & $3 \S{ABC}+3 \S{ABD}+3 \S{ACE}+\S{ABE}+\S{ACD}+\S{ADE}+\S{BCD}+\S{BCE}+\S{BDE}+\S{CDE} -\S{AD}-\S{AE}-\S{BC}-\S{DE}-\S{ABDE}-\S{ACDE}-2 \S{AB}-2 \S{AC}-2 \S{BD}-2 \S{CE}-2 \S{ABCD}-2 \S{ABCE} \geq 0$ & $-9 \tilde{S}_2 + 8\tilde{S}_3 \geq 0$ & yes \\ \hline
\end{tabular}
\end{table}

\clearpage
\newpage

\section{Extremal structure of the SQEC for $n\leq 5$}

The extreme rays of the SQEC are very simple and given by \eqref{eq:SQECrays}. In contrast, the extreme rays of its parent cone defined by all possible instances of SA and SSA turn out to be extremely complicated (e.g. there are millions of them for $n=5$). Hence we have decided to not include this data. Instead, we list here explicitly only the facets of the cone involving instances of SA and SSA and their symmetrizations, indicating which of the latter yield facets of the SQEC.

\begin{table}[h!]
\centering
$n=2$\\
\vspace{2mm}
\begin{tabular}{|r|c|c|c|}
\hline
$\#$ & SA+SSA & SQEC & facet? \\ \hline
1 & $\S{A} + \S{B} - \S{AB} \geq 0$ & $\tilde{S}_1 \geq 0$ & yes \\ \hline
\end{tabular}
\end{table}

\begin{table}[h!]
\centering
$n=3$\\
\vspace{2mm}
\begin{tabular}{|r|c|c|c|}
\hline
$\#$ & SA+SSA & SQEC & facet? \\ \hline
1 & $\S{A} + \S{B} - \S{AB} \geq 0$ & $2 \tilde{S}_1 - \tilde{S}_2 \geq 0$ & yes \\ \hline
2 & $\S{A} - \S{B} + \S{AB} \geq 0$ & $\tilde{S}_2 \geq 0$ &  \\ \hline
3 & $\S{AC} + \S{BC} + \S{C} - \S{ABC} \geq 0$ & $- \tilde{S}_1 + \tilde{S}_2 \geq 0$ & yes \\ \hline
\end{tabular}
\end{table}

\begin{table}[h!]
\centering
$n=4$\\
\vspace{2mm}
\begin{tabular}{|r|c|c|c|}
\hline
$\#$ & SA+SSA & SQEC & facet? \\ \hline
1 & $\S{A} + \S{B} - \S{AB} \geq 0$ & $2 \tilde{S}_1 - \tilde{S}_2 \geq 0$ & yes \\ \hline
2 & $\S{A} - \S{B} + \S{AB} \geq 0$ & $\tilde{S}_2 \geq 0$ &  \\ \hline
3 & $\S{A} + \S{BC} - \S{ABC} \geq 0$ & $\tilde{S}_1 \geq 0$ &  \\ \hline
4 & $-\S{A} + \S{BC} + \S{ABC} \geq 0$ & $- \tilde{S}_1 + 2 \tilde{S}_2 \geq 0$ &  \\ \hline
5 & $\S{AC} + \S{BC} - \S{C} - \S{ABC} \geq 0$ & $- \tilde{S}_1 + \tilde{S}_2 \geq 0$ & yes \\ \hline
\end{tabular}
\end{table}

\begin{table}[h!]
\centering
$n=5$\\
\vspace{2mm}
\begin{tabular}{|r|c|c|c|}
\hline
$\#$ & SA+SSA & SQEC & facet? \\ \hline
1 & $\S{A} + \S{B} - \S{AB} \geq 0$ & $2 \tilde{S}_1 - \tilde{S}_2 \geq 0$ & yes \\ \hline
2 & $\S{A} - \S{B} + \S{AB} \geq 0$ & $\tilde{S}_2 \geq 0$ &  \\ \hline
3 & $\S{A} + \S{BC} - \S{ABC} \geq 0$ & $\tilde{S}_1 + \tilde{S}_2 - \tilde{S}_3 \geq 0$ &  \\ \hline
4 & $\S{A} - \S{BC} + \S{ABC} \geq 0$ & $\tilde{S}_1 - \tilde{S}_2 + \tilde{S}_3 \geq 0$ &  \\ \hline
5 & $-\S{A} + \S{BC} + \S{ABC} \geq 0$ & $- \tilde{S}_1 + \tilde{S}_2 + \tilde{S}_3 \geq 0$ &  \\ \hline
6 & $-\S{C} +\S{BC} + \S{AC} - \S{ABC} \geq 0$ & $- \tilde{S}_1 + 2 \tilde{S}_2 - \tilde{S}_3 \geq 0$ & yes \\ \hline
7 & $-\S{A} -\S{B} + \S{AC} + \S{BC} \geq 0$ & $- \tilde{S}_1 + \tilde{S}_2 \geq 0$ &  \\ \hline
8 & $-\S{D} +\S{CD} + \S{ABD} - \S{ABCD} \geq 0$ & $- \tilde{S}_1 + \tilde{S}_3 \geq 0$ &  \\ \hline
9 & $-\S{CD} +\S{BCD} + \S{ACD} - \S{ABCD} \geq 0$ & $- \tilde{S}_2 + \tilde{S}_3 \geq 0$ & yes \\ \hline
\end{tabular}
\end{table}

\newpage
\clearpage

\section{Exact Volumes for Small $n$}\label{sec:volSmallN}

\begin{figure}[h!]
    \centering
    \includegraphics[width=10cm]{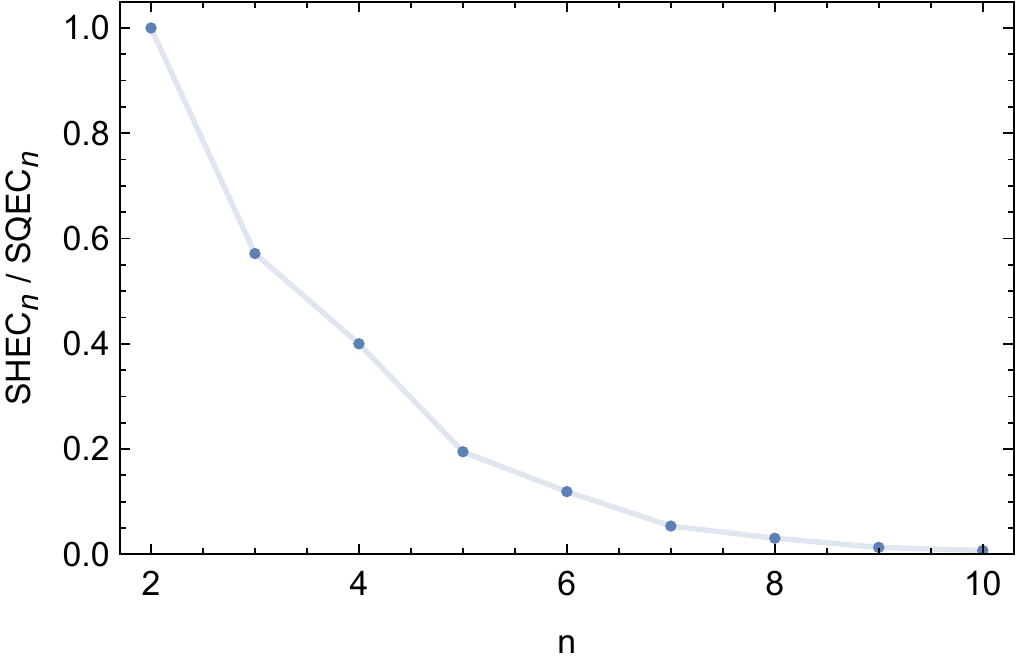}
    \caption{Exact ratio between the volumes of the SHEC$_n$ and SQEC$_n$ for $2\leq n \leq 10$.}
    \label{fig:volExact}
\end{figure}

From the extreme rays of the SHEC and SQEC we can compute the volume of these cones as explained in Section~\ref{sec:vols}. For small number of parties $n$, this volume can be easily computed exactly. In Fig.~\ref{fig:volExact}, and in the following table, we show the exact relative volume between the two cones for $n\leq 10$. Note that the SHEC becomes exponentially small compared to the SQEC as $n$ increases, see \eqref{eq:volRatio}.

\begin{table}[h!]
\begin{adjustwidth}{-2.1cm}{0cm}
    \centering
    \begin{tabular}{|c||c|c|c|c|c|c|c|c|c|}
        \hline
         $n$ & 2 & 3 & 4 & 5 & 6 & 7 & 8 & 9 & 10  \\
           \hline
           \hline
         $(\vol \text{SHEC}_n)^{-1}$ & 1 & 21 & 30 & 2772 & 4536 & 1127000 & 1982400 & 1036103250 & 1906410000 \\       \hline
         $(\vol \text{SQEC}_n)^{-1}$ & 1 & 12 & 12 & 540 & 540 & 60480 & 60480 & 13608000 & 13608000 \\ 
            \hline
         $\text{SHEC}_n / \text{SQEC}_n$ & 1 & 0.571 & 0.4 & 0.195 & 0.119 & 0.0537 & 0.0305 & 0.0131 & 0.00714 \\
         \hline
    \end{tabular}
    \label{tab:exact}
    \end{adjustwidth}
\end{table}

\newpage
\clearpage

\end{document}